\documentclass[runningheads]{llncs}
\usepackage{graphicx}
\usepackage{amsmath}
 \usepackage{enumerate}
 \usepackage{cite}
\begin{document}
\title{Super Resolution of Arterial Spin Labeling MR Imaging Using Unsupervised Multi-Scale Generative Adversarial Network}
%
%
\author{Jianan Cui\inst{1,2,\thanks{Contributed equally to this work}} \and
Kuang Gong\inst{2,3,*} \and
Paul Han \inst{3} \and
Huafeng Liu \inst{1} \and \\
Quanzheng Li \inst{2,3}}
\authorrunning{J. Cui et al.}
%
\institute{State Key Laboratory of Modern Optical Instrumentation, College of Optical Science and
	Engineering, Zhejiang University, Hangzhou, Zhejiang 310027 China 	\email{liuhf@zju.edu.cn}\\
	\and
	Center for Advanced Medical Computing and Analysis, Massachusetts General Hospital/Harvard Medical School, Boston, MA 02114 USA \and
Gordon Center for Medical Imaging, Massachusetts General Hospital/Harvard Medical School, Boston, MA 02114 USA \\
\email{Li.Quanzheng@mgh.harvard.edu} }

%
\maketitle              
\begin{abstract}
Arterial spin labeling (ASL) magnetic resonance imaging (MRI) is a powerful imaging technology that can measure cerebral blood flow (CBF) quantitatively. However, since only a small portion of blood is labeled compared to the whole tissue volume, conventional ASL suffers from low signal-to-noise ratio (SNR), poor spatial resolution, and long acquisition time. In this paper, we proposed a super-resolution method based on a multi-scale generative adversarial network (GAN) through unsupervised training. The network only needs the low-resolution (LR) ASL image itself for training and the T1-weighted image as the anatomical prior. No training pairs or pre-training are needed. A low-pass filter guided item was added as an additional loss to suppress the noise interference from the LR ASL image. After the network was trained, the super-resolution (SR) image was generated by supplying the upsampled LR ASL image and corresponding T1-weighted image to the generator of the last layer.  Performance of the proposed method was evaluated by comparing the peak signal-to-noise ratio (PSNR) and structural similarity index (SSIM) using normal-resolution (NR) ASL image (5.5 min acquisition) and high-resolution (HR) ASL image (44 min acquisition) as the ground truth. Compared to the nearest, linear, and spline interpolation methods, the proposed method recovers more detailed structure information, reduces the image noise visually, and achieves the highest PSNR and SSIM when using HR ASL image as the ground-truth.

\end{abstract}
\section{Introduction}
Arterial spin labeling (ASL) is a non-invasive magnetic resonance imaging (MRI) technique which uses water as a 
diffusible tracer to measure cerebral blood flow (CBF)\cite{williams1992magnetic}. Without a gadolinium-containing contrast agent, ASL is repeatable and can be applied to pediatric patients and patients with impaired renal function\cite{amukotuwa20163d,pedrosa2012arterial}. However, as the lifetime of the tracer is short---approximately the transport time from labeling position to the tissue---the data acquisition time is limited in a single-shot\cite{alsop2015recommended}.  Thus, ASL images usually suffer from low signal-to-noise ratio (SNR) and limited spatial resolution. 

Due to the great potential of ASL in clinical and research applications, developing advanced acquisition sequence and processing methods to achieve high-resolution, high-SNR ASL has been a very active research area. Regarding post-processing methods, non-local mean \cite{coupe2008optimized} and temporal filtering \cite{petr2010improving} has been developed to improve the SNR of ASL without additional training data.  Partial volume correction methods \cite{chappell2011partial,liang2013improved,meuree2019patch} have also been proposed to improve the perfusion signal from ASL images with limited spatial resolutions. As for reconstruction approaches, parallel imaging \cite{boland2018accelerated}, compressed sensing \cite{han2016whole}, spatial \cite{petr2010denoising} or spatio-temporal constraints \cite{spann2017spatio,fang2015spatio} have been proposed to recover high-SNR ASL images from noisy or sparsely sampled raw data.

Recently, deep learning has achieved great success in computer vision tasks when large number of training data exist. It has been applied to ASL image denoising and showed better results than state-of-the-arts methods\cite{gong2017boosting,kim2018improving,xie2020denoising,ulas2018deepasl,gong2019arterial}. Specifically, Zheng et al proposed a two-stage multi-loss super resolution (SR) network that can both improve spatial resolution and SNR of the ASL images\cite{li2019two}. These prior arts mostly focused on supervised deep learning where a large number of high-quality ASL images are required. However, it is not easy to acquire such kind of training labels in clinical practice, especially for high-resolution ASL, due to the MR sequence limit and long acquisition time. 

Previously, we have proposed an unsupervised deep learning framework for PET image denoising\cite{cui1,cui2,cui3}. In this paper, we explored the possibility of using unsupervised Generative Adversarial Network (GAN) to perform ASL super-resolution. It does not require high-resolution ASL images as training labels and no extra training is needed. The main contributions of this work include:
\begin{enumerate}[(1)]
 \item The low-resolution ASL image itself was used as the training label and only one single ASL volume was needed for GAN training. No training labels nor large number of datasets is needed. Registered T1-weighted image was fed to the network as a separate channel to provide anatomical prior information. 
\item Inspired by the pyramid structure of the sinGAN structure \cite{shaham2019singan}  which extracted features from multi-scales, our network consisted of multi-scale feature-extraction layers. After network training, generator of the last layer containing the finest scale information was used to generate the final super-resolution ASL image.
\item Considering the noisy character of the ASL images, a super low-pass filter loss term was added as an additional loss item to reduce the image noise and stabilize the network training.
\item The voxel size of the generated SR image can be adjusted at will as there is no requirement of zoom-in ratio and the proposed method is fully unsupervised.
\end{enumerate}

\section{Method}
\begin{figure}
	\includegraphics[width=\textwidth]{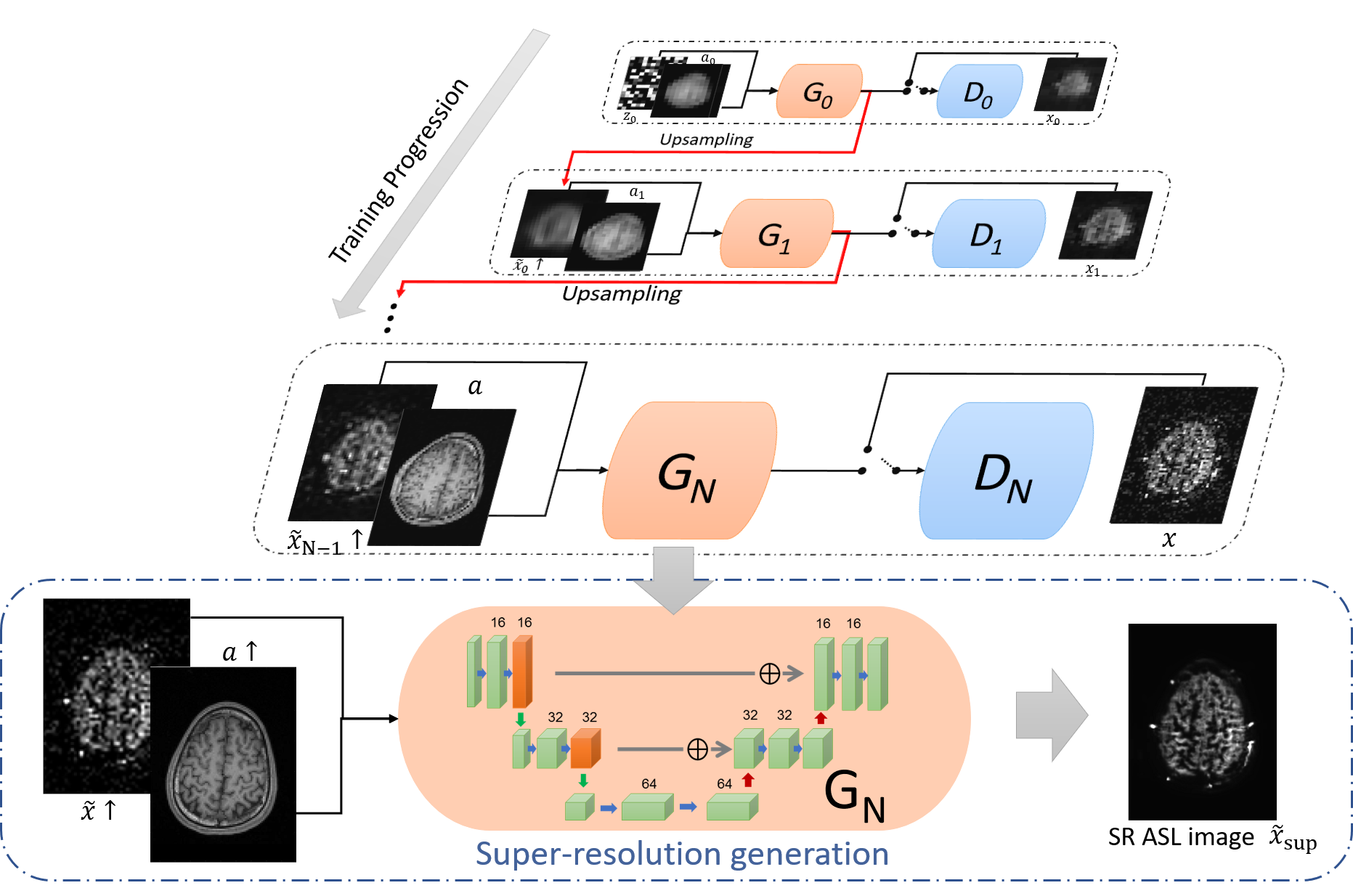}
	\caption{Diagram of the proposed framework. The network contains multi-scale layers. Each layer includes a generator and a discriminator.} \label{fig1}
\end{figure}

Diagram of the whole generative framework is shown in Fig. \ref{fig1}. It can be performed in two steps: multi-scale training and super-resolution generation. 

\subsection{Multi-Scale Training Step} A pyramid of generators $\left\{G_0,...,G_N\right\}$ was trained against an ASL image pyramid of $x:\left\{x_0,...x_N\right\}$ layer by layer progressively. Meanwhile, a T1-weighted image pyramid of $a:\left\{a_0,...,a_N\right\}$ was inserted to the generators in another channel to provide anatomical information. The T1 image $a$ was coregistered with the low resolution (LR) ASL image $x$. $x_n$ and $a_n$ are downsampled from $x$ and $a$ by a factor $r^{N-n}$. At the very beginning, spatial white Gaussian noise $z_0$ was injected as the initial input of generator $G_0$. Then each generator accepted the upsampled generative result $\widetilde{x}_{n-1}{\uparrow}$ from the upper layer as the input to generate image $\widetilde{x}_{n}$ in a higher spatial resolution,
\begin{equation}
\widetilde{x}_{n} = \left\{
\begin{matrix}
G_0\left ( z_0,a_0 \right ) & \ \ \ \ n=0 \\
G_n\left ( {\widetilde{x}_{n-1}}{\uparrow},a_n \right )   & \ \ \ \   0 < n\leq N 
\end{matrix}\right. .
\end{equation}

For each layer's training, the generator $G_n$ intended to produce realistic image like the training label $x_n$ to fool the associated discriminator $D_n$, while the discriminator was trained to distinguish the generated image $G_n\left ( {\widetilde{x}_{n-1}}{\uparrow},a_n \right )$ from real $x_n$. All the generators and the discriminators share the same architectures which are a three-layer 3D Unet\cite{cciccek20163d} and Markovian discriminator\cite{isola2017image,li2016precomputed}, respectively.
Each generator used the upper-layer generator's parameters as the initial parameters. The network was trained by the Adam optimizer with a learning rate of ${10^{-3}}$. 2000 epochs were run for each layer.

\subsection{Super-Resolution Generation Step} 
The LR ASL image $x$ was upsampled to $x{\uparrow}$ and $a{\uparrow}$ is T1-weighted image of the same size.  Both $x{\uparrow}$ and $a{\uparrow}$ were supplied to the last generator $G_N$ and the output of the generator $G_N$ is the final SR ASL image. The multi-scale architecture has been verified by \cite{shaham2019singan}, which can recover structure features from the coarsest scale to the finest scale. After training, the generator $G_N$ learned fine details and texture information from the last layer. Based on it, we can recover the missing structure details of the upsampled image $x{\uparrow}$.
\begin{equation}
\widetilde{x}_{sup} = G_N\left ( {x}{\uparrow},a{\uparrow} \right )  .
\end{equation}

\subsection{Loss Function}
Each GAN in the network was trained with the loss function which is combined by three part: adversarial loss, mean squared error (MSE) loss and low-pass filter loss,
\begin{equation}
\min_{G_n} \max_{D_n}L_{adv}\left ( G_n,D_n \right )+ \alpha L_{mse}\left ( G_n \right )+\beta L_{lp}\left ( G_n \right ),
\end{equation} 
where $\alpha$ and $\beta$ are the weighted parameters for MSE loss and Gaussian filter loss, respectively. The adversarial loss was WGAN-GP loss\cite{gulrajani2017improved}. The MSE loss aims to improve Peak SNR (PSNR),
\begin{equation}
L_{mse}= {\left \| G_n\left ( {\widetilde{x}_{n-1}}{\uparrow},a_n \right )-x_n \right \|}^{2}.
\end{equation}
The low-pass filter loss term calculates the MSE between the generated image and the real image label after they pass a super low-pass filter $F$. This term can reduce the influence of noise and it is formulated as:
\begin{equation}
L_{lp}= {\left \| F(G_n\left ( {\widetilde{x}_{n-1}}{\uparrow},a_n \right ))-F(x_n )\right \|}^{2}.
\end{equation}
In this paper, we use Gaussian filter with a sigma of 5.

\section{Experiment}
\subsection{Overall Design}
Two experiments are designed to evaluate the performance of our proposed method. Firstly, we used low-resolution (LR) ASL image as the training label and generated normal-resolution (NR) ASL image. As the LR ASL image was downsampled from NR ASL image, we can calculate peak SNR (PSNR) and structural similarity index (SSIM) for quantitative analysis. In the second experiment, we tried to generate SR image which has the same voxel size as the T1-weighted image by training NR ASL image.  The second experiment can better show the practical usability of the proposed method. 
\subsection{Data Acquisition and Generation}
The ASL MRI image was acquired from a healthy volunteer by a 3T whole-body scanner (Magnetom Tim Trio, Siemens Healthcare, Erlangen, Germany) using pseudo-continuous ASL (pCASL) with bSSFP readout (total labeling duration: 1500 ms; post-labeling delay time: 1.2s). The matrix size of the ASL MRI image is $128\times96\times48$ with a voxel size of $1.875\times1.875\times2.5 mm^{3}$ and TR/TE = 3.93/1.73ms. There are ASL images of two different spatial resolution considering different acquisition time: 44 min for high-resolution ASL image and 5.5 min for normal-resolution ASL image. A three-plane localizer and a T1-weighted magnetization-prepared rapid gradient echo (MPRAGE) were performed after the ASL acquisition. The matrix size of the T1-weighted image is $224\times176\times256$ with a voxel size of $0.9766\times0.9766\times1 mm^{3}$.

In order to verify the performance of the proposed method quantitatively, we generated the LR ASL image (matrix size: $64\times48\times48$; voxel size: $3.75\times3.75\times2.5 mm^{3}$) by downsampling the NR ASL image (5.5 min acquisition). As the T1-weighted image was paired with the upsampled ASL image during the train step and the super-resolution generation step, the T1 image was registered to different scales of the ASL image by Advanced Normalization Tools (ANTs) \cite{avants2009advanced}. 
 
\section{Results}
\begin{figure}[!htp]
	\centering
	\includegraphics[height=3.8in]{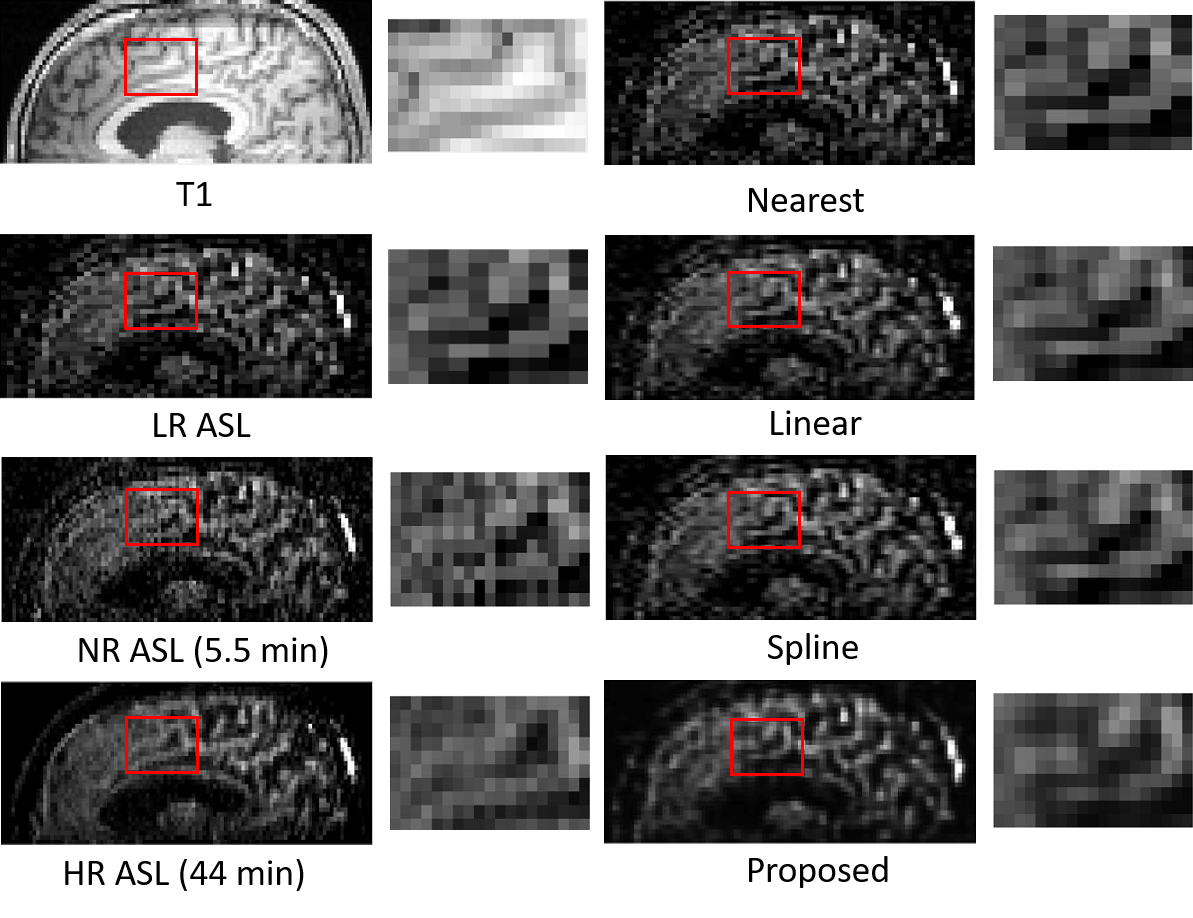}
	\caption{Visual comparison (sagital view) of different reference images and different super-resolution method results. The super-resolution was performed on LR ASL image in the first row. } \label{fig3}
\end{figure}

\begin{figure}[!htp]
	\centering
	\includegraphics[trim=0.cm 0cm 0cm 0cm, clip, width=4.5in,height=4.5in]{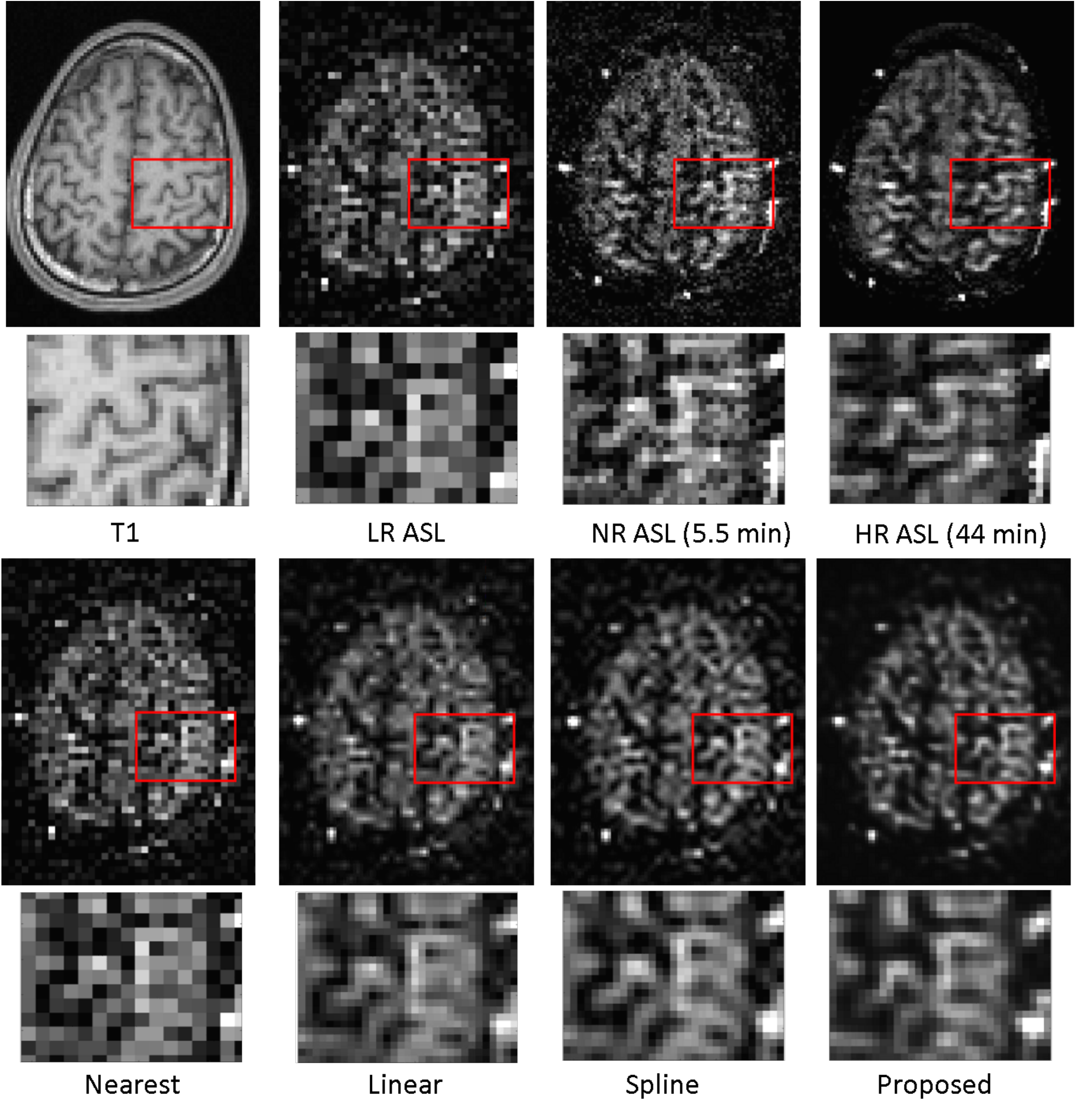}
	\caption{Visual comparison (transaxial view) of different reference images and different super-resolution method results. The super-resolution was performed on LR ASL image in the first row. } \label{fig2}
\end{figure}

In the first experiment, the network was trained using the LR ASL image as the training label. The super-resolution results were compared with the nearest interpolation, linear interpolation, and spline interpolation results, as shown in Fig. \ref{fig3} (sagital view) and Fig. \ref{fig2} (transaxial view). It can be seen that images of the proposed method have a good visual appearance with clear boundaries and low image noise. The quantitative results of PSNR and SSIM are shown in Table.~\ref{tab1}. One interesting observation is that when using NR ASL image as the reference image, both linear and spline interpolation have better PSNR and SSIM than the proposed method. However, the proposed method achieves the highest PSNR and SSIM when using 44-min HR ASL image as the reference image. As the NR ASL image is still noisy and does not have as many details as the HR ASL image, this conflicting conclusion actually proves that the proposed method does utilize the structure information from T1-weighted image which can not be observed from the NR ASL image. 

\begin{table}
	\caption{Quantitative results using NR ASL and HR ASL as reference}\label{tab1}
	\centering
	\begin{tabular}{|c|c|c|c|c|}
		\hline
		groundtruth&\multicolumn{2}{|c|}{NR ASL}&\multicolumn{2}{|c|}{HR ASL}\\
		\hline
		&  PSNR & SSIM&  PSNR & SSIM\\
		\hline
		Nearest & 31.5249& 0.7507 & 30.9289 & 0.5320\\
		Linear  & {\bfseries 33.6745} &{\bfseries0.8158} &32.9041 & 0.5851\\
		Spline  & 32.8720& 0.7816 & 32.1518 & 0.5536\\
		Proposed & 32.5795&0.7795&{\bfseries 33.3723}  & {\bfseries 0.6488}\\
		\hline
	\end{tabular}
\end{table}

\begin{figure}[h]
	\centering
	\includegraphics[width=4.5in]{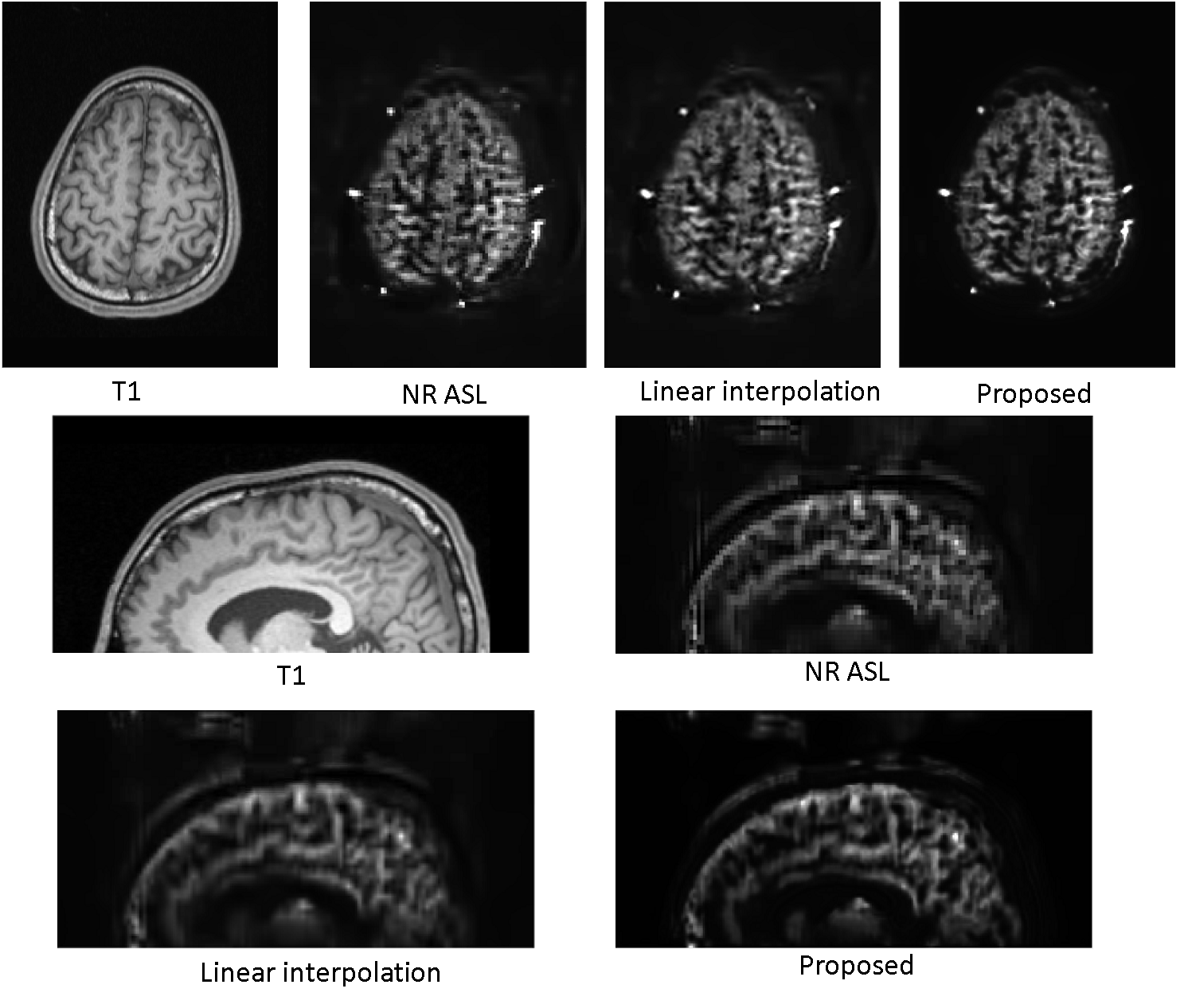}
	\caption{Visual comparison of different reference images and different super-resolution method results. The super-resolution was performed on NR ASL image in the first row.} \label{fig4}
\end{figure}

For the second experiment, we trained the network by using the NR ASL image as the training label (Fig. \ref{fig4}), and then generated SR ASL image with the same voxel size ($0.9766\times0.9766\times1 mm^{3}$) as the T1-weighted image. Successful super-resolution generation of this experiment can prove that this proposed framework can be applied to any existing ASL protocols by translating the voxel size and resolution to that of the T1-weighted image. There is no limit on the voxel-size ratio between the original ASL and the T1-weighted image, which is quite flexible. The results in transaxial view and sagital view shown in Fig. \ref{fig4} demonstrates this point. The linear interpolation result was chosen for comparison as it has higher PSNR and SSIM than the nearest and spline interpolation in the quantification. The structure of the proposed result is sharp and clear while the linear interpolation result is a little bit blurred.

\section{Conclusion}
In this paper, we proposed an unsupervised multi-scale GAN framework for ASL image super-resolution. Corresponding T1-weighted image inserted into the network works as an anatomical prior to guide the ASL super-resolution generation process. After being trained layer by layer progressively, the last-layer generator can capture the finest feature to produce realistic super-resolution images. In-vivo results show that the proposed framework can simultaneously improve spatial resolution while also reducing the image noise with the help of low-pass-filter loss term. Our future work will focus on more quantitative analysis of the generated CBF map as well as more data evaluations. 

\section{Acknowledgement}
This work is supported in part by the National Key Technology Research and Development Program of China (No: 2017YFE0104000, 2016YFC1300302), and by the National Natural Science Foundation of China (No: U1809204, 61525106, 81873908, 61701436).
%
%

%

\end{document}